\documentclass[symmetry,article,publish,moreauthors,latex]{mdpi}
\UseRawInputEncoding
\usepackage{amsmath}
\usepackage{amsfonts}
\usepackage{amssymb,bm}
\usepackage{siunitx}
\usepackage{color}
\usepackage{braket}
\usepackage{graphics}

\firstpage{1}
\pubvolume{15}
\issuenum{5}
\articlenumber{1011}
\pubyear{2023}
\copyrightyear{2023}
\externaleditor{Academic Editor: {Vladimir Dobrev}} 
\datereceived{{18 April 2023}}
\dateaccepted{{28 April 2023}}
\datepublished{{1 May 2023}}
\hreflink{https://doi.org/10.3390/sym15051011} 

\Title{Comparison of the Lifshitz Theory Using the Nonconventional Fit of Response
Functions with Precise Measurements of the Casimir Force}

\TitleCitation{Comparison of the Lifshitz Theory Using the Nonconventional Fit of Response
Functions with Precise Measurements of the Casimir Force}

\Author{
Galina~L.~Klimchitskaya ${}^{1,2}$\orcidA{} and
Vladimir~M.~Mostepanenko ${}^{1,2,3,}$*\orcidA{}}

\AuthorNames{
Galina L. Klimchitskaya, Vladimir M. Mostepanenko}

\AuthorCitation{Klimchitskaya, G.L.; Mostepanenko, V.M.}

\address{%
${}^{1}$ \quad Central Astronomical Observatory at Pulkovo of the Russian Academy of Sciences, 196140 Saint Petersburg, Russia; g.klimchitskaya@gmail.com\\
${}^{2}$ \quad Peter the Great Saint Petersburg
Polytechnic University, 195251 Saint Petersburg, Russia\\

$^{3}$ \quad Kazan Federal University, 420008 Kazan, Russia}

\corres{Correspondence: vmostepa@gmail.com}

\abstract{
It is known that the fundamental Lifshitz theory, which is based on the first
principles of thermal quantum field theory, experiences difficulties when
compared with precise measurements of the Casimir force.
We analyzed the nonconventional fit of the response functions of many materials
along the imaginary frequency axis to the empirical model of ``modified''
oscillators, which was recently proposed in the literature. According to our
results, this model is unacceptable because at high frequencies it leads
to the asymptotic behavior of the response functions, which is in
contradiction with that following from the fundamental physical
principles. We calculated the Casimir interaction in the configurations
of several precise experiments using the Lifshitz theory and the
response functions to the quantized electromagnetic field expressed in terms of
modified oscillators and demonstrated that the obtained results are excluded
by the measurement data. This invalidated a claim made in the literature that the
Casimir--van der Waals forces calculated using these response functions are
in remarkable agreement with the experimental values. Possible reasons
for a disagreement between experiment and theory are discussed, and the
way to improve the situation is indicated.}


\begin{document}

\newcommand{\ve}{\varepsilon}
\newcommand{\vIe}{\varepsilon_{\vphantom{\int}\rm I}}
\newcommand{\vMe}{\varepsilon_{\vphantom{\int}\rm M}}
\section{Introduction}

The physical nature of the Casimir (van der Waals in the nonrelativistic limit) forces
determined by the zero-point and thermal fluctuations of the quantized electromagnetic
field is widely covered in the literature (see the monographs \cite{1,2,3,4,5,6,7,8,9}
and the references therein). The fundamental theoretical description of these forces,
which is based on thermal quantum field theory in the Matsubara formulation, was
developed by E.\ M.\ Lifshitz \cite{10,11}. It expresses the force value as a functional
of the frequency-dependent response functions (dielectric permittivities) of the interacting
bodies defined along the imaginary frequency axis. In the last few years, many
experiments on measuring the Casimir--van der Waals force were performed,
and their results were compared with theory (see the reviews
\cite{12,13,14,15}). For symmetric configurations (e.g., for two parallel plates
or a sphere above a plate), the Casimir force acts perpendicular to the surfaces.
If, however, the rotational symmetry is violated (e.g., for the plates covered with
longitudinal coaxial corrugations), the lateral Casimir force may arise as well \cite{8}.

Currently, the Casimir--van der Waals forces are not only actively investigated
in elementary particle physics, gravitation, cosmology, fundamental atomic,
molecular, and condensed matter physics \cite{1,2,3,4,5,6,7,8,9}, but find
application in micromechanics, microelectronics and nanoelectronics including
organic electronics, various sensors, microswitches, analog integrated
circuits, analog-to-digital and thermoelectric converters, etc. (see, e.g.,
\cite{16,17,18,19,20,21,22,23,24,25,26,27,28,29,30,31,32,32a}). Therefore,
the possibility to reliably predict the force value on the basis of the fundamental
Lifshitz theory is of the utmost importance. This, however, turned out to
be a challenge. The point is that, to perform reliable computations using the
Lifshitz theory, one needs to have at hand the response functions of the
interacting bodies to the quantized electromagnetic field over rather wide
frequency regions. Even when this information seems to be available,
it happens that the computational results are in conflict with the measurement
data \cite{8,12,15,33,34}.

Recently \cite{35}, the ``self-consistent'' (by the authors' terminology)
response functions for 55 materials over the full region of imaginary
frequencies necessary for the calculation of Casimir--van der Waals
forces were determined including common metals, organic and inorganic
semiconductors, and insulators. According to \cite{35}, the
Casimir--van der Waals forces calculated using these response functions ``are
in remarkable agreement with the experimental values reported over the
span of the past half-century.''

The imaginary parts of the response functions of materials studied in
\cite{35} were compiled from the measured complex indices of refraction
in different experiments. Thereafter, the response functions along the
imaginary frequency axis have been obtained using the principle of
causality expressed in the form of the Kramers--Kronig
relations, which is the generally recognized way of calculations
\cite{6,7,8,9,12,13,14,15}. However, the article \cite{35} claimed that the obtained
response functions along the imaginary frequency axis are better fit not
to the conventional Lorentz oscillators \cite{5,8,36,37,38}, but to the
model of nonconventional ``modified'' oscillators. The latter model predicts
that at high frequencies $\omega$ the response functions approach unity
within the term decreasing as $\omega^{-\alpha}$, where $0<\alpha<2$, in
place of the standard term decreasing as $\omega^{-2}$.

In this paper, we call the reader's attention to the fact that the
high-frequency behavior of the response functions suggested by the model
of modified oscillators is in contradiction with fundamental physical principles.
Using the data of the most precise experiments on measuring the Casimir
interaction between metallic surfaces, we also show that theoretical
predictions of the Lifshitz theory found with the model of modified
oscillators are excluded at a high confidence level. Thus, the conclusion
of \cite{35} about a remarkable agreement between the theoretical
results obtained using the ``self-consistent'' response functions and the
experimental values is invalidated. The reason why an unjustified
conclusion has been made is that \cite{35} did not use the data of
the most precise experiments when performing the theory--experiment
comparison.

The paper is organized as follows. In Section \ref{sec2}, we confront the
representations of response functions using the models of the Lorentz
and modified oscillators. Section \ref{sec3} contains a comparison of the
theoretical results computed with the model of modified oscillators and
the measured Casimir interaction between Au surfaces. In Section \ref{sec4}, a
similar comparison is performed with the measured Casimir interaction
between magnetic (Ni) surfaces. In Section \ref{sec5}, the reader will find our
discussion of possible reasons for a disagreement between experiment
and theory, and in Section \ref{sec6} we give our conclusions, where a line of attack
on the problem is directed.

\section{Representation of the Dielectric Functions Using the Lorentz
Oscillators and the Nonconventional Modified Oscillators\label{sec2}}

It is common knowledge that a reasonably accurate and yet simple representation
for the response functions of different insulator materials to the electromagnetic field
is given by the sum of the appropriate number of Lorentz oscillators describing the bound
(core) electrons \cite{5,8,36,37,38}:
\begin{equation}
\vIe(\omega)=1+\sum_{j=1}^{K}\frac{g_j}{1-\left(\frac{\omega}{\omega_j}\right)^2
-i\frac{\gamma_j\omega}{\omega_j^2}},
\label{eq1}
\end{equation}
\noindent
where $K$ is the number of oscillators, $\omega_j\neq 0$ are the oscillator
frequencies, $\gamma_j$ are the relaxation parameters, and $g_j$ are proportional
to the oscillator strengths. The representation (\ref{eq1}) works well for the
materials whose molecules do not possess intrinsic dipole moments.
For the polar dielectrics, the molecules of which possess intrinsic dipole moments
oriented in an electromagnetic field, one more (Debye) term should be added on
the right-hand side of~(\ref{eq1}).

If the boundary material is metallic, there is also an additional oscillator term
in (\ref{eq1}) with zero oscillator frequency, $\omega_0=0$, describing the
conduction (free) electrons. As a result, the response functions of metallic materials
take the form:
\begin{equation}
\vMe(\omega)=-\frac{g_0}{\omega(\omega+i\gamma_0)}+\vIe(\omega),
\label{eq2}
\end{equation}
\noindent
which is known as the dielectric permittivity of the Drude model. In this case, $g_0$
has the meaning of the plasma frequency squared, $g_0=\omega_p^2$.

The response functions (\ref{eq1}) and (\ref{eq2}) satisfy the principle of causality,
which is expressed mathematically in the form of the Kramers--Kronig
relations connecting their imaginary and real parts \cite{39}. Substituting
$\omega=i\xi$, one obtains the response functions of insulators and metals
defined along the imaginary frequency axis:
\begin{eqnarray}
&&
\vIe(i\xi)=1+\sum_{j=1}^{K}\frac{g_j}{1+\left(\frac{\xi}{\omega_j}\right)^2
+\frac{\gamma_j\xi}{\omega_j^2}},
\nonumber \\
&&
\vMe(i\xi)=\frac{\omega_p^2}{\xi(\xi+\gamma_0)}+\vIe(i\xi).
\label{eq3}
\end{eqnarray}
\noindent
Note that the same expressions result from ${\rm Im}\ve_{\vphantom{\int}\rm I,M}$
by using the dispersion relation \cite{39}:

\begin{equation}
\ve_{\vphantom{\int}\rm I,M}(i\xi)=1+\frac{2}{\pi}\int_{0}^{\infty}
\frac{\omega\,{\rm Im}\ve_{\vphantom{\int}\rm I,M}(\omega)}{\omega^2+\xi^2}\,d\omega.
\label{eq4}
\end{equation}

It should be pointed out that at short separations between the interacting bodies
the major contribution to the Lifshitz formula for the Casimir--van der Waals force
is given by the large $\xi\gg\gamma_j$. Because of this, (\ref{eq3}) can be
rewritten in a simpler form:
\begin{eqnarray}
&&
\vIe(i\xi)\approx1+\sum_{j=1}^{K}\frac{g_j}{1+\left(\frac{\xi}{\omega_j}\right)^2},
\nonumber \\
&&
\vMe(i\xi)\approx \vIe(i\xi),
\label{eq5}
\end{eqnarray}
\noindent
which is referred to as the Ninham--Parsegian representation \cite{36,37}.
Thus, at short separations, the force is mostly determined by the core (bound)
electrons.

As to the case of large separations, where the major contribution to the force
is given by the zero Matsubara frequency (see the next section), one arrives at
\begin{eqnarray}
&&
\vIe(i\xi)\approx\vIe(0)=1+\sum_{j=1}^{K}{g_j},
\nonumber \\
&&
\vMe(i\xi)\approx\frac{\omega_p^2}{\xi(\xi+\gamma_0)}.
\label{eq6}
\end{eqnarray}
\noindent
This means that for metals the force value is determined by the free (conduction)
electrons.

For materials that exhibit an electronic polarization only (for Si, for instance),
the sum in (\ref{eq5}) may be replaced by one effective oscillator term with
a frequency $\omega_{\vphantom{\int}\rm UV}$ belonging to the ultraviolet region.
As for materials that also exhibit an ionic polarization (SiO$_2$, for instance),
their response function may be presented as a sum of one effective oscillator with
a frequency $\omega_{\vphantom{\int}\rm UV}$ and another one whose frequency
$\omega_{\vphantom{\int}\rm IR}$ belongs to the infrared region:
\begin{equation}
\vIe(i\xi)=1+\frac{g_{\vphantom{\int}\rm UV}}{1+\left(
\frac{\xi}{\omega_{\vphantom{\int}\rm UV}}\right)^2} +
\frac{g_{\vphantom{\int}\rm IR}}{1+\left(
\frac{\xi}{\omega_{\vphantom{\int}\rm IR}}\right)^2}.
\label{eq7}
\end{equation}

Equations (\ref{eq3}), (\ref{eq5}), and (\ref{eq7})
have been extensively used
in the literature (see, e.g., \cite{38,40}) to fit the measured data for
the response functions of many materials. The obtained expressions were
extensively applied to compute the Casimir--van der Waals
forces (see the lists of references in the monographs \cite{2,3,4,5,6,7,8,9}).

At high frequencies, (\ref{eq3}), (\ref{eq5}), and (\ref{eq7}) predict
the following similar asymptotic behavior for the response functions of both
insulating and metallic materials:
\begin{equation}
\ve(\omega)-1\sim\frac{1}{\omega^2},\qquad\ve(i\xi)-1\sim\frac{1}{\xi^2}.
\label{eq8}
\end{equation}

According to \cite{35}, however, this prediction is incorrect. In support
of this claim, some optical data for water, SiO$_2$, and LiF were collected,
which seemingly yielded the electronic contribution to the dielectric functions along
the imaginary frequency axis obeying at high frequencies the law \cite{35}:
\begin{equation}
\ve_{el}(i\xi)-1\sim \frac{1}{\xi^{\alpha}},
\label{eq9}
\end{equation}
\noindent
where $0<\alpha<2$ in place of (\ref{eq8}).

On this basis, Reference \cite{35} argued that the electronic contribution to the
response function commonly described by the first two terms on the right-hand
side of (\ref{eq7}) can be modified to an empirical relationship:
\begin{equation}
\ve_{el}(i\xi)=1+\frac{g_{\vphantom{\int}\rm UV}}{1+\left(
\frac{\xi}{\omega_{\vphantom{\int}\rm UV}}\right)^{\alpha}}.
\label{eq10}
\end{equation}

It is easily seen, however, that (\ref{eq9}) and, thus, (\ref{eq10})
are invalid. As explained in \cite{39}, if the field frequency $\omega$
is much larger compared to the frequencies of almost all atomic electrons, the
latter can be considered as free particles not interacting between themselves
and with atomic nuclei. The velocities of electrons in atoms $v$ are small
compared to the speed of light $c$. Because of this, the distances
$2\pi v/\omega$ traveled by electrons during a period of the electromagnetic
wave are small in comparison to the wavelength $2\pi c/\omega$.

As a consequence, when finding the velocity of an electron in the
electromagnetic field of
a wave, the latter can be considered as spatially homogeneous. Under these
conditions, solving the equation of motion and calculating the polarization
of the material by summing over all electrons in the unit volume, one
arrives at \cite{39}
\begin{equation}
\ve(\omega)=1-\frac{4\pi Ne^2}{m_e\omega^2},\quad
\ve(i\xi)=1+\frac{4\pi Ne^2}{m_e\xi^2},
\label{eq11}
\end{equation}
\noindent
where $m_e$ and $e$ are the mass and charge of an electron and $N$ is the
number of electrons in all atoms of the unit of volume of a material.

We underline that (\ref{eq11}) is a universal result following from the
basic physical principles, and it is valid for the response functions of
any material: insulator, metal, or semiconductor. For light elements, the
application region of (\ref{eq11}) starts from the far ultraviolet
(6~eV--10~eV) and for heavier elements from the X-ray
frequencies (100~eV--100~keV) \cite{39}. Thus, if some used sets of the
measured optical data (often taken from different sources) lead to the
response function $\ve(i\xi)$, which does not satisfy (\ref{eq11})
at sufficiently high frequencies, one should cast doubt on these data.

\section{Comparison Between Measured and Calculated Casimir--van~der~Waals Forces
Using the Nonconventional Fit to the Modified Oscillators for Gold Test Bodies\label{sec3}}

According to \cite{35}, the ``self-consistent'' response function of Au
presented by (\ref{eq2}) using the fit to a modified oscillator is given by
(see the Supplemental Materials in \cite{41} to \cite{35})
\begin{equation}
\ve_{\rm Au}(i\xi)=1+\frac{(9.1)^2}{\xi(\xi+0.06)}+
\frac{6.5}{1+\left(\frac{\xi}{5.9}\right)^{1.42}},
\label{eq12}
\end{equation}
\noindent
where $\omega_p=9.1~$eV, $\gamma_0=0.06~$eV, $g_{\vphantom{\int}\rm UV}=6.5$,
$\omega_{\vphantom{\int}\rm UV}=5.9~$eV, and $\alpha=1.42$ (the frequency
$\xi$ is measured in eV).

Reference \cite{35} claimed that the theoretical Casimir--van der Waals
forces calculated using the ``self-consistent'' response functions
in the framework of the Lifshitz theory match remarkably well with the
experimentally measured forces. It is well known that the most precise
measurements of the Casimir--van der Waals forces were performed
between an Au-coated sphere and an Au-coated plate separated with a
vacuum gap \cite{8,12}. In Figure 5 of \cite{35},
the theory--experiment comparison of this kind was presented,
however, only for
two and not the most precise experiments \cite{42,43} performed long ago.

Thus, in the experiment \cite{42}, the Casimir force was measured
between the Au-coated surfaces of a sphere and a plate by means of an
atomic force microscope operated in static mode. Here, we calculated
the Casimir force in the experimental configuration of \cite{42}
using the Lifshitz theory and the ``self-consistent'' response function
(\ref{eq12}) with the properly accounted-for effect of surface roughness as
described in \cite{44,45}.

The Casimir force acting between a sphere of radius $R$ spaced at a
distance $a \ll R$ above a plate at temperature $T$ in thermal equilibrium with
the environment is expressed by the following Lifshitz formula \cite{8,12}:
\begin{equation}
{F}(a)=k_BTR
\sum_{l=0}^{\infty}{\vphantom{\sum}}^{\prime}\int_0^{\infty}\!\!k_{\bot} dk_{\bot}
\sum_{\alpha}\ln\left[1-r_{\alpha}^2(i\xi_l,k_{\bot})e^{-2aq_l}\right].
\label{eq13}
\end{equation}
\noindent
Here, $k_B$ is the Boltzmann constant, $k_{\bot}$ is the magnitude of the wave vector
component along the plate, $\xi_l = 2\pi k_BTl/\hbar$ are the Matsubara frequencies with
$l = 0,1,2,...,$ the prime on the first summation sign divides the term with $l = 0$ by 2, and
the summation in $\alpha$ is over two independent polarizations of the electromagnetic
field, transverse magnetic ($\alpha$ = TM) and transverse electric ($\alpha$ = TE).
The reflection coefficients calculated at the pure imaginary Matsubara frequencies
are given by
\begin{equation}
r_{\rm TM}(i\xi_l,k_{\bot})=\frac{\varepsilon_lq_l - k_l}{\varepsilon_lq_l + k_l}, \quad
r_{\rm TE}(i\xi_l,k_{\bot})=\frac{\mu_lq_l - k_l}{\mu_lq_l + k_l},
\label{eq14}
\end{equation}
\noindent
where $\varepsilon_l = \varepsilon(i\xi_l)$, the magnetic permeability $\mu_l = \mu(i\xi_l)$
describes the response of materials to the magnetic field (for Au $\mu(\omega) = 1$),
and the following notations are introduced:
\begin{equation}
q_l=\left(k_{\bot}^2+\frac{\xi_l^2}{c^2}\right)^{1/2}, \quad
k_l=\left[k_{\bot}^2+\varepsilon_l\mu_l\frac{\xi_l^2}{c^2}\right]^{1/2}.
\label{eq15}
\end{equation}
\noindent
Equation (\ref{eq13}) is written in the proximity force approximation \cite{8,12}.
The exact theory using the scattering approach demonstrates that, at short separations
$a \ll R$, the corrections to~(\ref{eq13}) do not exceed the fraction of $a/R$
\cite{46,47,48,49,50,51,52,53}. These corrections are negligibly small in the
configurations of precise experiments on measuring the Casimir force \cite{8,12,33,34}.

The theoretical results taking into account the surface roughness are obtained from~
(\ref{eq13}) by the geometrical averaging over the measured roughness profiles of
the plate and sphere surfaces \cite{6,12}:
\begin{equation}
F_R(a)=\sum_{i,k}v_i^{(1)}v_k^{(2)}F(a+H_0^{(1)}+H_0^{(2)}-h_i^{(1)}-h_k^{(2)}),
\label{eq16}
\end{equation}
\noindent
where $v_i^{(1)} (v_k^{(2)})$ are the fractions of the plate (sphere) areas with heights
$h_i^{(1)} (h_k^{(2)})$. Here, $H_0^{(1,2)}$ are the zero levels relative to which the mean
values of the roughness profiles on both bodies are zero. It was shown \cite{8,12,33}
that at short separations, where the effect of surface roughness should be taken into
account, the more fundamental analysis based on the scattering approach \cite{54}
leads to approximately the same results as the method of geometrical averaging.

\begin{figure}[!h]
\vspace*{-3.5cm}
\centerline{\hspace*{-1cm}
\includegraphics[width=3.5in]{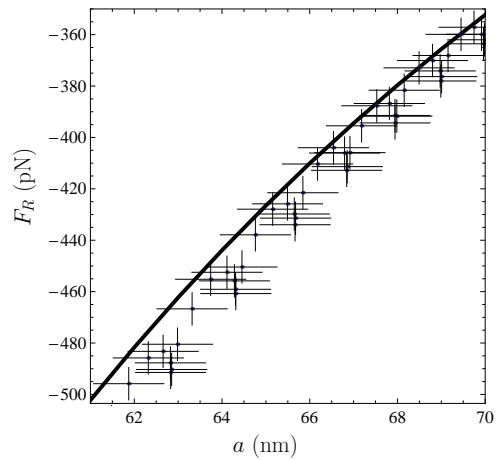}}
\vspace*{-2.7cm}
\caption{\label{fg1}Predictions of the Lifshitz theory for the Casimir force
between the Au surfaces of a sphere and a plate obtained using the
modified oscillator in the experimental configuration of \cite{42}
are shown by the black band. The measurement data of \cite{42,44}
are indicated as crosses, whose arms are determined at the 95\%
confidence level. }
\end{figure}

The computational results obtained using (\ref{eq12}), (\ref{eq13}), and (\ref{eq16})
are shown in Figure~\ref{fg1} by the black band, whose width is determined by the
theoretical errors calculated at the 95\% confidence level. In the same
figure, the measurement data are indicated as crosses, whose arms show
the experimental errors in measuring the absolute separations and
forces, which were also determined at the 95\% confidence level.

As is seen in Figure~\ref{fg1}, the agreement between experiment and theory
cannot be called remarkably good because some of the crosses do not even touch
the theoretical band. Reference \cite{35} recognized the presence of
some deviations between the theoretical results computed using the
modified oscillators and the measurement data, especially at short
separations, but attributes them to the probable role of surface
roughness and to the subnanometer errors in the estimation of the
separation distances. However, in the theory--experiment comparison
presented here in Figure~\ref{fg1}, both of these effects were addressed
quantitatively and taken into account.

\begin{figure}[!t]
\vspace*{-3.5cm}
\centerline{\hspace*{-2cm}
\includegraphics[width=6.8in]{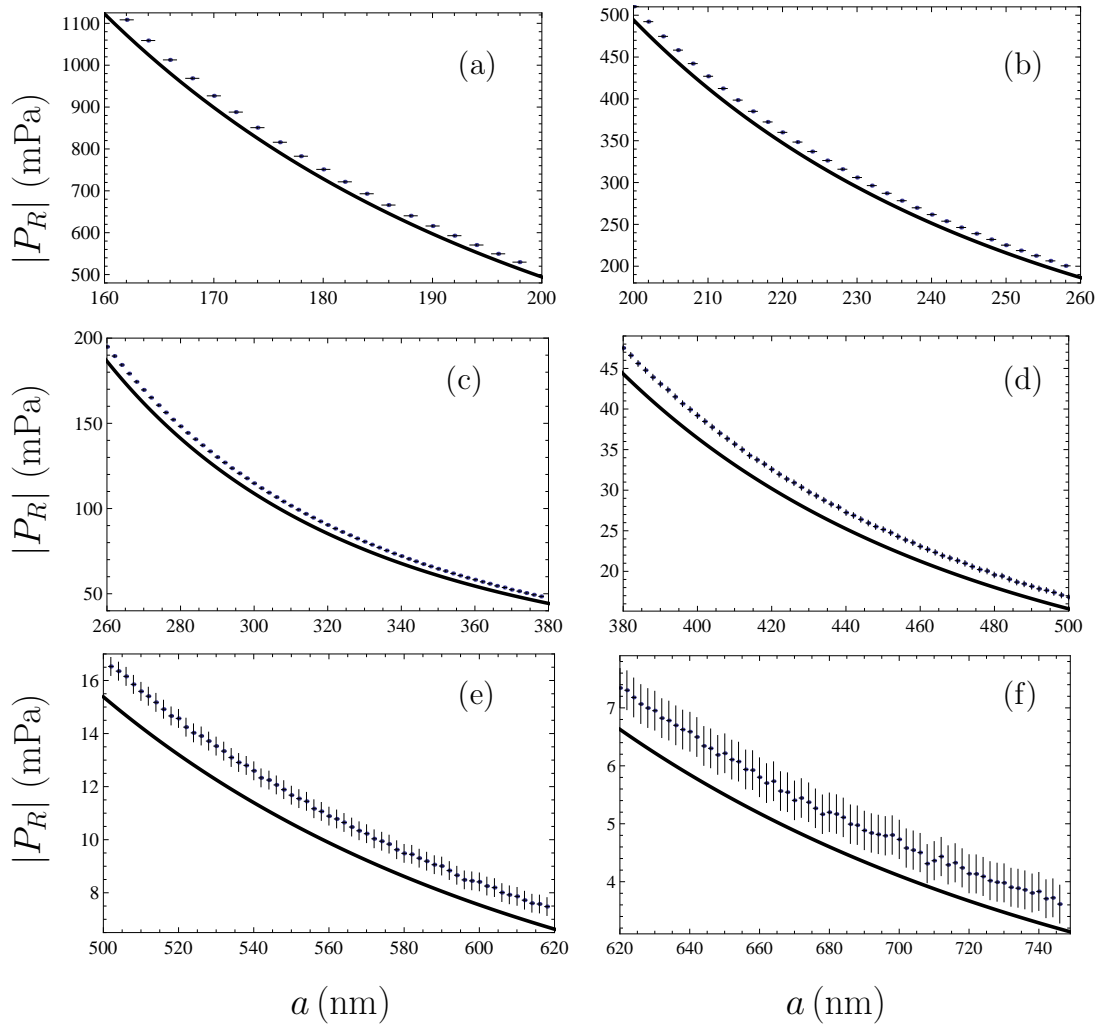}}
\vspace*{-6.5cm}
\caption{\label{fg2} Predictions of the Lifshitz theory for the effective Casimir
pressure between two Au plates obtained using the modified oscillator
in the experimental configuration of \cite{45,46} are shown by the
black bands over different separation intervals. The measurement data of
\cite{45,46} are indicated as crosses, whose arms are determined at
the 95\% confidence level. In all subfigures a-f, the theoretical predictions
obtained using (12) are excluded by the measurement data.}
\end{figure}

The more precise modern experiments on measuring the Casimir
interaction demonstrated the total discrepancy between the theoretical
predictions using the dielectric functions of \cite{35} and the
measurement data. In Figure~\ref{fg2}a--f, the experimental data of
\cite{55,56} for determining the effective Casimir pressure $P_R(a)$
between the Au-coated surfaces of two parallel plates by means of a
micromechanical torsional oscillator are shown as crosses over six
different separation intervals.
The theoretical bands are computed
here using the dielectric function (12) with the account of the surface
roughness by the Lifshitz formula \cite{8,12}:
\begin{equation}
P_R(a)=-\frac{1}{2\pi R}\frac{dF_R(a)}{da}.
\label{eq17}
\end{equation}
\noindent
Both the arms of the crosses and
the widths of the bands are again found at the 95\% confidence level
(see \cite{55,56} for the details of the computations).

As is seen in Figure~\ref{fg2}, the theoretical predictions found by using the
``self-consistent'' dielectric functions of Au (12) are excluded by
the measurement data at the 95\% confidence level over the entire
measurement range.

\begin{figure}[!t]
\vspace*{-.5cm}
\centerline{\hspace*{-2.3cm}
\includegraphics[width=5in]{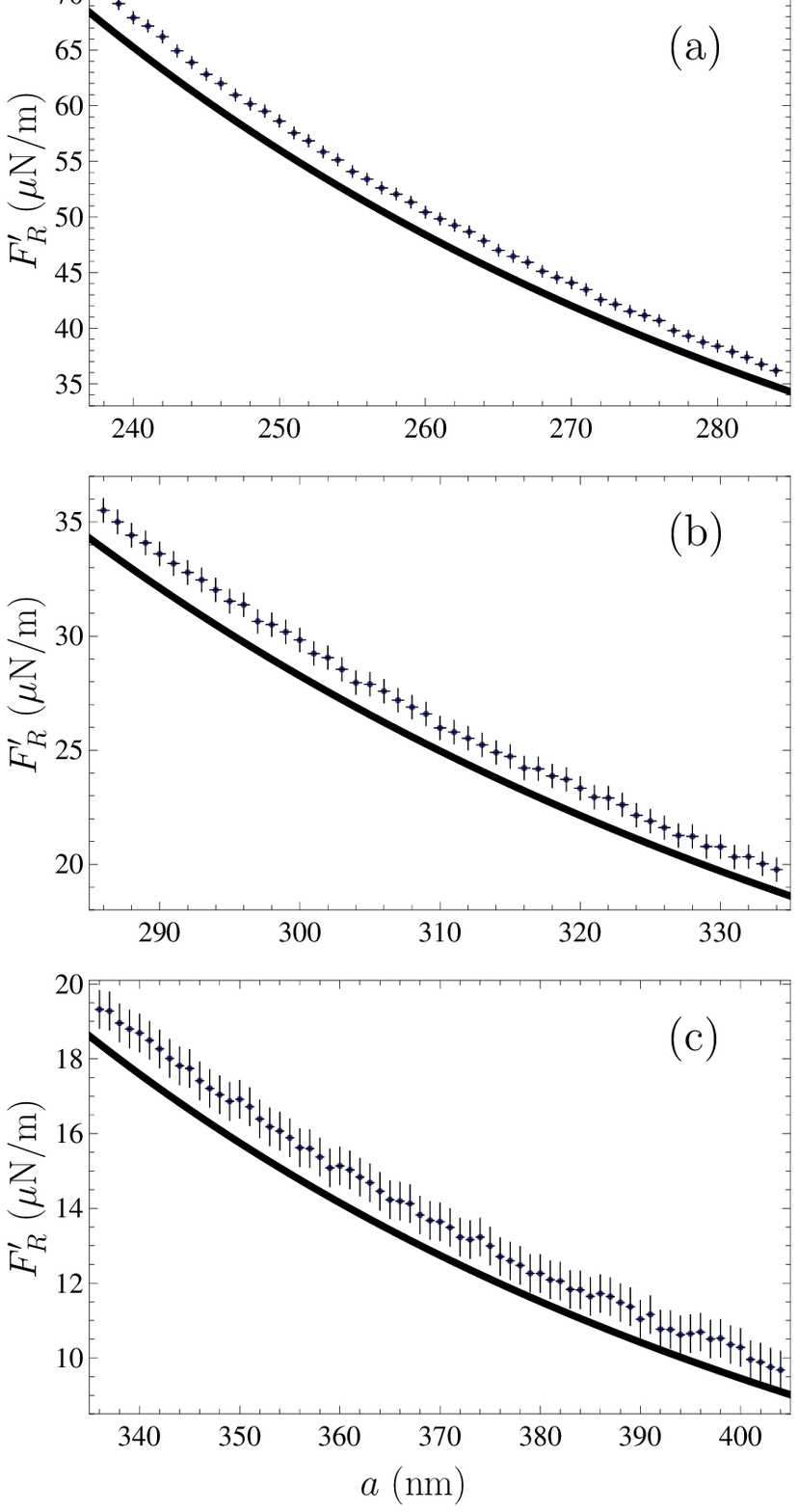}}
\vspace*{-6.5cm}
\caption{\label{fg3} Predictions of the Lifshitz theory for the gradient of the
Casimir force between the Au surfaces of a sphere and a plate obtained
using the modified oscillator in the experimental configuration of
\cite{47} are shown by the black bands over different separation
intervals. The measurement data of~\cite{47} are indicated as
crosses, whose arms are determined at the 67\% confidence level. In all
subfigures a-c, the theoretical predictions obtained using (12) are excluded by
the measurement data.}
\end{figure}

In one more precise experiment, the gradient of the Casimir force $dF_R(a)/da$
acting between the Au-coated surfaces of a sphere and a plate was
measured by means of a dynamic atomic force microscope \cite{57}.
The experimental data measured in this experiment are shown as
crosses over the three separation intervals in Figure~\ref{fg3}a--c.

The
theoretical bands are again computed with the account of the surface
roughness by using (\ref{eq13}), (\ref{eq16}), and the ``self-consistent'' response
function (\ref{eq12}). In this case, all errors were determined at the 67\% confidence
level. From Figure~\ref{fg3}, one can conclude that the theoretical predictions
obtained using the response function (\ref{eq12}) are excluded by the
measurement data of \cite{57}.

We note that the data of the most precise experiments \cite{55,56,57},
measuring the Casimir interaction between two Au surfaces separated by a
vacuum gap, were not considered and compared with the suggested
approach of \cite{35} using a
nonconventional fit of the response functions to the modified
oscillators. As a result, an invalid conclusion has
been made that the Casimir--van der Waals interaction computed using
these response functions agree remarkably well with the experimental
results.

\section{Comparison Between Measured and Calculated Casimir--van~der~Waals Forces
Using the Nonconventional Fit to the Modified Oscillators for Nickel Test Bodies\label{sec4}}

Measurements of the Casimir interaction for the test bodies made of a magnetic metal
Ni play an especially important role in Casimir physics because the magnetic
properties of a material, along with the dielectric ones, produce a pronounced
impact on the force value. The gradient of the Casimir force between the Ni-coated
surfaces of a sphere and a plate separated by a vacuum gap was measured in
\cite{58,59} by means of a dynamic atomic force microscope.

Here, we compared the obtained measurement data with the theoretical predictions
of the Lifshitz theory found using the ``self-consistent'' response function of Ni
presented in the form of a modified oscillator in \cite{35,41}:
\begin{equation}
\ve_{\rm Ni}(i\xi)=1+\frac{(4.33)^2}{\xi(\xi+0.0195)}+
\frac{115}{1+\left(\frac{\xi}{0.61}\right)^{1.35}},
\label{eq18}
\end{equation}
\noindent
where $\omega_p=4.33~$eV, $\gamma_0=0.0195~$eV, $g_{\vphantom{\int}\rm UV}=115$,
$\omega_{\vphantom{\int}\rm UV}=0.61~$eV, and $\alpha=1.35$.

\begin{figure}[!h]
\vspace*{-2.5cm}
\centerline{\hspace*{-0.5cm}
\includegraphics[width=5.5in]{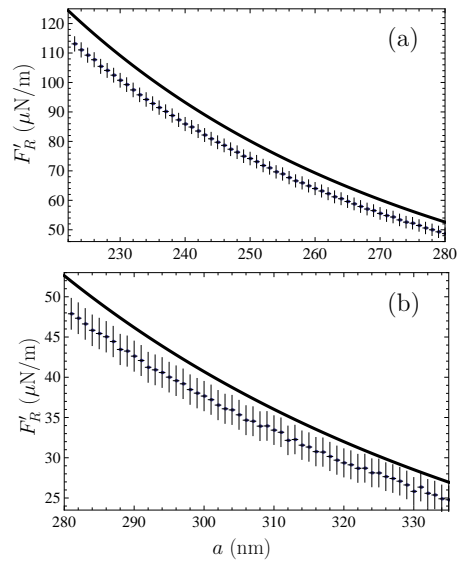}}
\vspace*{-9.3cm}
\caption{\label{fg4} Predictions of the Lifshitz theory for the gradient of the
Casimir force between the Ni surfaces of a sphere and a plate obtained
using the modified oscillator in the experimental configuration of
\cite{48,49} are shown by the black bands over different separation
intervals. The measurement data of \cite{48,49} are indicated as
crosses, whose arms are determined at the 95\% confidence level. In both subfigures a,b,
the theoretical predictions obtained using (18) are excluded by the measurement data.}
\end{figure}

The theoretical band for the gradient of the Casimir force calculated using
(\ref{eq13}) and (\ref{eq18}) with the account of the surface roughness is shown in Figure~\ref{fg4}a,b
in two different separation intervals. These calculations take into account the
static magnetic permeability of Ni, $\mu_{0}(0)=110$, entering the Lifshitz formula
through the reflection coefficients (\ref{eq14}) at zero Matsubara frequency \cite{58,59}.
Note that the magnetic properties of a material do not contribute to all Matsubara terms of
(\ref{eq13}) with $l \geq 1$ because $\mu(i\xi)$ quickly drops to unity with increasing
$\xi$ \cite{60}.
The experimental data for the measured gradients are indicated as crosses. Here, we
recalculated the arms of these crosses to the 95\% confidence level (in the original
publications, the errors were determined at the 67\% confidence level).

As is seen in Figure~\ref{fg4}, the theoretical predictions obtained using the
nonconventional fit of the optical data to the modified oscillator (\ref{eq18}) are
excluded by the measurement data over the separation distances from 222 to 335~nm.

A comparison between Figure~\ref{fg4} and Figure~\ref{fg3} shows an important difference
between the cases of magnetic (Ni) and nonmagnetic (Au) metals. {}From Figure~\ref{fg4},
it is seen that for a magnetic metal the theoretically predicted force gradients are
larger than the measured values, whereas for a nonmagnetic metal the measured force
gradients are in excess of the computed results. This fact was used in \cite{58,59}
to underline the existence of a serious unresolved problem in the calculation of the Casimir--van der Waals
forces. Although in the experiments mentioned above a gap between the computed and
observed values is of about a few percent, in measuring the differential Casimir forces
the theoretical predictions of the Lifshitz theory differ from the measured values by
up to a factor of 1000 \cite{61}. This is a reason why the importance of the observed
disagreement between experiment and theory should not be underestimated.

\section{Discussion\label{sec5}}

In the foregoing, we analyzed the recently proposed nonconventional
fit of the response functions of a wide class of materials along the
imaginary frequency axis to the so-called ``modified'' oscillators
\cite{35,41}. It was demonstrated that the mere form of the modified
oscillator is unacceptable because it leads to an incorrect asymptotic
behavior of the response functions at high frequencies, which is in
contradiction with the fundamental physical principles. What is more, we
calculated the Casimir--van der Waals interaction in the configurations
of several precise experiments using the formalism of the modified
oscillators and found that the obtained computational results are
excluded by the measurement data.

The opposite result obtained in \cite{35} is explained by the
fact that the authors did not consider the most precise experiments
and presented the results of their comparison in the logarithmic scale,
which does not allow an informative discrimination between different
lines. Note also that \cite{35} arrived at a misleading result
that ``In the case of metals,... the role of interband transitions on
the magnitude of van der Waals-Casimir forces becomes crucial once the
ratio of charge carriers to total electrons in the systems becomes small.''
In fact, as outlined in Section 2, for a particular metal, the relative role
of the core and conduction electrons depends on the separation distance
between the test bodies made of this metal. At separations much smaller
than a micrometer, the major contribution to the Casimir--van der Waals
force is given by the core (bound) electrons, whereas at separations in
excess of several micrometers the force value is determined by the
conduction electrons. Keeping in mind the wide application areas of the
Casimir--van der Waals forces and the many publications using the
dielectric functions of diverse materials in the computations, the above
clarifications regarding \cite{35} seem pertinent.

\section{Conclusions\label{sec6}}

We conclude with a short remark on the general situation in the theory
of Casimir--van der Waals forces. It has been long known that the
standard description of free charge carriers by means of the Drude
model leads to contradictions between experiment and theoretical
Casimir forces even if the conventional fit of the optical data
to the Lorentz oscillators is employed
\cite{8,12,14,15,55,56,57,58,59,61}. When using the standard sets of
optical data from Palik's handbook \cite{62}, the theory comes to
an agreement with the measurement results if the conduction electrons
are described by the dissipationless plasma model
\cite{8,12,14,15,55,56,57,58,59,61}. This fact, however, has no commonly
accepted theoretical explanation (note that with the values of the
plasma frequency found in \cite{35} an agreement between experiment
and theory is lacking regardless of what model is chosen for a
description of conduction electrons). Similar problems arise for the
Casimir force acting between insulator test bodies \cite{8,12, 63,64,65,66}.
There is a possibility to deal with these problems using the so-called ``weighted''
Kramers--Kronig relations \cite{67}, which allow the mathematical derivation of the
response function of a material outside the region where it is measured with
sufficient precision. This method, however, relies on the known behavior
at zero frequency, which is different for the Drude and plasma response functions.

It was hypothesized that the above problems derive from the fact that the Drude
model is not applicable in the area of the low-frequency
s-polarized evanescent waves, where it has no sufficient experimental
confirmation \cite{68,69}. An experiment was proposed allowing one to check
this hypothesis \cite{68,69}. The possible resolution of the problem might
be found in the search for spatially nonlocal generalizations of the
Drude model at low frequencies \cite{70,71}. Several promising
opportunities on this way are in sight.

\vspace{6pt}

\funding{
The work of G.L.K. was partially funded by the Ministry of Science and Higher
Education of the Russian Federation (``The World-Class Research Center:
Advanced Digital Technologies'', Contract No. 075-15-2022-311 dated
April 20, 2022).
The research of V.M.M. was partially carried out in accordance with the Strategic
Academic Leadership Program ``Priority 2030'' of the Kazan Federal University.
}

\end{paracol}

\reftitle{References}


\end{document}